\documentstyle[12pt]{article}
\begin{document}

\author{R. Vilela Mendes \\
Grupo de F\'\i sica-Matem\'atica, Complexo II, Universidade de Lisboa,\\
Av. Gama Pinto 2, 1699 Lisboa Codex, Portugal \and J. A. Dente \\
Laborat\'orio de Mecatr\'onica, DEEC, Instituto Superior T\'ecnico,\\
Av. Rovisco Pais, 1096 Lisboa Codex, Portugal}
\title{Boundary-layer control by electric fields: A feasibility study }
\maketitle

\begin{abstract}
A problem of great concern in aviation and submarine propulsion is the
control of the boundary layer and, in particular, the methods to extend the
laminar region as a means to decrease noise and fuel consumption. In this
paper we study the flow of air along an airfoil when a layer of ionized gas
and a longitudinal electric field are created in the boundary layer region.
By deriving scaling solutions and more accurate numerical solutions we
discuss the possibility of achieving significant boundary layer control for
realistic physical parameters. Practical design formulas and criteria are
obtained. We also discuss the perspectives for active control of the
laminar-to-turbulent transition fluctuations by electromagnetic field
modulation.
\end{abstract}

\section{Boundary layers and boundary layer control}

Whether a flow is laminar or turbulent, the effects of the viscosity of the
fluid are greatest in regions close to solid boundaries. The region close to
the boundary, in which the velocity varies from zero, relative to the
surface, up to its full value, is called the {\it boundary layer}. The
concept of boundary layer introduced by Prandtl in 1904, was a most
significant advance in fluid dynamics, in the sense that it simplified the
study by separating the flow in two parts: (1) the region where velocity
gradients are large enough to produce appreciable viscous forces - the
boundary layer itself - and (2) the external region where viscous forces are
negligible compared to other forces. From the computational point of view
the concept of boundary layer also plays a significant role because, rather
than having to deal with time-consuming general purpose finite-element
codes, results of comparable precision may be obtained by fast and
relatively simple finite-difference implicit algorithms.

When a fluid flows past a solid body, an airfoil for example, a {\it laminar
boundary layer} develops, in general only for a very small distance near the
leading edge, followed by a transition to a {\it turbulent boundary layer}.
Nevertheless, because near the solid wall velocity fluctuations must die
out, below the turbulent region there always is a {\it laminar sub-layer}
which in general is very small (of the order of a micrometer). The
transition from the laminar to the turbulent region is controlled by the 
{\it local Reynolds number,} defined in terms of the effective thickness of
the boundary layer. It also depends on the smoothness of the surface and on
the external perturbations. The {\it skin friction drag} is proportional to
the gradient of the longitudinal velocity at the solid boundary. Because of
the mixing properties of the turbulent layer, the gradient in the laminar
sub-layer is much greater than the gradient at a fully laminar layer.
Therefore transition from a laminar to a turbulent layer greatly increases
the skin friction drag. Another effect to be taken into account is the {\it %
separation} of the boundary layer, which occurs at points where the pressure
gradient along the surface reverses sign. The eddies, generated by the
resulting reverse flow, disturb the flow and form a {\it wake} where energy
dissipation decreases the pressure, thereby increasing the {\it pressure drag%
}.

Because of the very large ratio between laminar and turbulent skin friction
drag, much effort has been devoted to develop techniques to delay the
transition as a means of decreasing fuel consumption and noise. Care should
however be taken because, in general, a turbulent boundary layer is more
stable towards separation than a laminar boundary layer. Some of the active
control techniques that have been proposed include suction of slow-moving
fluid through slots or a porous surface, use of compliant walls and wall
cooling (or wall heating for liquids). Injection of fast-moving fluid, on
the other hand, is effective in avoiding separation but it increases
turbulence. Most of these ideas are fairly old (see for example \cite
{Schlichting},\cite{Young}) however, in view of their interest for the
applications and to obtain a more accurate characterization of the physical
mechanisms, studies of boundary layer control using these aerodynamic
methods are still, at present, being vigorously pursued (see for example 
\cite{ElHady} \cite{Joslin} \cite{Antonia} \cite{Lucey} and papers in \cite
{Bushnell})

Another class of techniques for active boundary layer control consists in
acting on the flow by means of electromagnetic forces. Here different
techniques should be envisaged according to whether the fluid is weakly
conducting (an electrolyte like seawater or an ionized gas) or a good
conductor (like a liquid metal). Proposals for boundary layer control by
electromagnetic forces are also relatively old and trace its origin at least
to the papers of Gailitis and Lielausis\cite{Gailitis}, Tsinober and Shtern%
\cite{Tsinober1} and Moffat\cite{Moffat} in the sixties. Interest in these
techniques has revived in recent years and some more accurate calculations
and experimental verifications have been carried out, mostly in the context
of electrolyte fluids\cite{Tsinober2} \cite{Henoch}.

In this paper we will be concerned with the flow of air along an airfoil
when a layer of ionized gas is created on the boundary layer region. Local
ionization of the air along the airfoil is not practical from the
technological point of view, therefore we will assume that a stream of
ionized air (or some other ionized gas) is injected through a backwards
facing slot placed slightly behind the stagnation point (Fig.1). The body
force that we consider to be acting in the ionized fluid is a longitudinal
(along the flow) electric field created by a series of plate electrodes
transversal to the flow and placed inside the airfoil with the edges on the
airfoil surface.

The emphasis of our study is on finding physically reasonable ranges of
parameters and analytic approximations that might lead to simple designing
procedures. For this purpose, before the numerical calculation of Section 3,
we dedicate some time to the study of scaling solutions and analytical
approximations.

The provisional conclusions of our study are that it is possible to use this
technique to control the profile of the boundary layer laminar region. With
the rates of ionization that are needed and the injection method, it is
probably unrealistic to expect that the laminar region may be extended over
all the airfoil in normal (aviation) working conditions. Therefore this
method should be used in conjunction with methods for control of turbulent
boundary layers (riblets, large-eddy breakups, additives, etc.) in the rear
part of the airfoil. Also the injection of the stream of ionized gas in the
leading edge may, by increasing the velocity component normal to the
airfoil, create turbulence. Therefore it seems advisable to have a
compensating suction region after the injection slot. The ionization rate
will also improve if the gas extracted through the suction region is
recycled through the ionizer.

Notice also that, once the fluid in the boundary layer is ionized, large
scale velocity fluctuations may be detected by a few local probes. This
raises the possibility of obtaining a negative feedback effect by an
appropriate time-dependent modulation of the electric field. By controlling
the growth of the velocity fluctuation in the transition region, a further
extension of the laminar region may be obtained. This is briefly discussed
in the last section of the paper.

The overall conclusion is that, when used in conjunction with other
techniques, as explained above, the method of boundary layer control by
electric fields might be interesting from the fuel consumption point of
view. This study was carried out as a preparation for an experiment being
set up in our Mechatronics Laboratory.

\section{Ionized boundary layers with electric fields}

\subsection{The boundary layer equations}

We use orthogonal curvilinear coordinates with $\widetilde{x}$ parallel to
the surface along the flow and $\widetilde{y}$ normal to the surface. If $%
\kappa \delta $ is small ($\kappa $ denoting the curvature and $\delta $ the
boundary layer thickness) the conservation and momentum equations in the
incompressible fluid approximation may be written 
\begin{equation}
\label{2.1a}\frac{\partial \widetilde{u}}{\partial \widetilde{x}}+\frac{%
\partial \widetilde{v}}{\partial \widetilde{y}}=0 
\end{equation}
\begin{equation}
\label{2.1b}\frac{\partial \widetilde{u}}{\partial \widetilde{t}}+\widetilde{%
u}\frac{\partial \widetilde{u}}{\partial \widetilde{x}}+\widetilde{v}\frac{%
\partial \widetilde{u}}{\partial \widetilde{y}}=-\frac 1{\widetilde{\rho }_m}%
\frac{\partial \widetilde{p}}{\partial \widetilde{x}}+\widetilde{\nu }\left( 
\frac{\partial ^2\widetilde{u}}{\partial \widetilde{x}^2}+\frac{\partial ^2%
\widetilde{u}}{\partial \widetilde{y}^2}\right) +\frac 1{\widetilde{\rho }_m}%
\widetilde{\sigma }_e(\widetilde{x},\widetilde{y})\widetilde{E}_x(\widetilde{%
x},\widetilde{y}) 
\end{equation}
\begin{equation}
\label{2.1c}\frac{\partial \widetilde{v}}{\partial \widetilde{t}}+\widetilde{%
u}\frac{\partial \widetilde{v}}{\partial \widetilde{x}}+\widetilde{v}\frac{%
\partial \widetilde{v}}{\partial \widetilde{y}}=-\frac 1{\widetilde{\rho }_m}%
\frac{\partial \widetilde{p}}{\partial \widetilde{y}}+\widetilde{\nu }\left( 
\frac{\partial ^2\widetilde{v}}{\partial \widetilde{x}^2}+\frac{\partial ^2%
\widetilde{v}}{\partial \widetilde{y}^2}\right) +\frac 1{\widetilde{\rho }_m}%
\widetilde{\sigma }_e(\widetilde{x},\widetilde{y})\widetilde{E}_y(\widetilde{%
x},\widetilde{y}) 
\end{equation}
$\widetilde{u}$ and $\widetilde{v}$ are the components of the fluid velocity
field along the $\widetilde{x}$ and $\widetilde{y}$ directions. $\widetilde{%
\rho }_m$ is the mass density, $\widetilde{\sigma }_e$ the electric charge
density and $\widetilde{E}$ an applied electric field. The tilde denotes
quantities in physical dimensions to be distinguished from the adimensional
quantities defined below. We consider typical values $L_r$, $\delta _r$, $%
U_r $, $\rho _r$,$\nu _r$,$\sigma _r$, $E_r$ as reference values for,
respectively, the airfoil width, the boundary layer thickness, the fluid
velocity, the fluid mass density, the kinematic viscosity, the fluid charge
density and the electric field. Then we define the adimensional quantities 
\begin{equation}
\label{2.2a}t=\widetilde{t}\frac{U_r}{L_r}\bigskip\ ,\bigskip\ x=\frac{%
\widetilde{x}}{L_r}\bigskip\ ,\bigskip\ y=\frac{\widetilde{y}}{\delta _r}%
\bigskip\ ,\bigskip\ u=\frac{\widetilde{u}}{U_r}\bigskip\ ,\bigskip\ v=\frac{%
\widetilde{v}L_r}{U_r\delta _r} 
\end{equation}
\begin{equation}
\label{2.2b}\rho _m=\frac{\widetilde{\rho }_m}{\rho _r}\bigskip\ ,\bigskip\
p=\frac{\widetilde{p}}{\rho _rU_r^2}\bigskip\ ,\bigskip R_L=\frac{U_rL_r}{%
\nu _r} 
\end{equation}
\begin{equation}
\label{2.2c}\nu =\frac{\widetilde{\nu }}{\nu _r}\bigskip\ ,\bigskip\ \sigma =%
\frac{\widetilde{\sigma }}{\sigma _r}\bigskip\ ,\bigskip\ E=\frac{\widetilde{%
E}}{E_r} 
\end{equation}
In general $R_L>>1$. Neglecting terms of order $\frac 1{R_L}$ and $\frac{%
\delta _r^2}{L_r^2}$ we obtain, for stationary solutions $\left( \frac{%
\partial u}{\partial t}=\frac{\partial v}{\partial t}=0\right) $ 
\begin{equation}
\label{2.3a}\frac{\partial u}{\partial x}+\frac{\partial v}{\partial y}=0 
\end{equation}
\begin{equation}
\label{2.3b}u\frac{\partial u}{\partial x}+v\frac{\partial u}{\partial y}%
=-\frac 1{\rho _m}\frac{\partial p}{\partial x}+\nu \omega \frac{\partial ^2u%
}{\partial y^2}+\gamma \frac 1{\rho _m}\sigma (x,y)E_x(x,y) 
\end{equation}
\begin{equation}
\label{2.3c}\frac{\partial p}{\partial y}=\frac{\delta _r}{L_r}\gamma \sigma
(x,y)E_y(x,y) 
\end{equation}
where $\omega =\frac{L_r^2}{\delta _r^2R_L}=\frac{L_r\nu _r}{\delta _r^2U_r}$
and $\gamma =\frac{L_r\sigma _rE_r}{U_r^2\rho _r}$. Unless the electric
field component normal to the airfoil is very large, one has $\frac{\partial
p}{\partial y}\approx 0$ and the pressure term in the second equation may be
expressed in terms of the fluid velocity $u_e$ far away from the airfoil 
\begin{equation}
\label{2.4}u\frac{\partial u}{\partial x}+v\frac{\partial u}{\partial y}=u_e%
\frac{\partial u_e}{\partial x}+\nu \omega \frac{\partial ^2u}{\partial y^2}%
+\gamma \frac 1{\rho _m}\sigma (x,y)E_x(x,y) 
\end{equation}
To take into account turbulence effects one should also replace in (\ref{2.4}%
) the velocity fields $u$ and $v$ by $u+u^{^{\prime }}$ and $v+v^{^{\prime
}} $, $u^{^{\prime }}$ and $v^{^{\prime }}$ being fluctuation fields with
zero mean, $\overline{u^{^{\prime }}}=0$, $\overline{v^{^{\prime }}}=0$. The
effect of the turbulent field on the mean flow is now obtained by taking
mean values. In a two-dimensional turbulent boundary layer the dominant eddy
stress is $\overline{-u^{^{\prime }}v^{^{\prime }}}$. Assuming the eddy
shear stress $\overline{-u^{^{\prime }}v^{^{\prime }}}$ and the mean rate of
strain $\frac{\partial u}{\partial y}$ to be linearly related 
\begin{equation}
\label{2.5}\overline{-u^{^{\prime }}v^{^{\prime }}}=\epsilon \frac{\partial u%
}{\partial y} 
\end{equation}
one obtains finally 
\begin{equation}
\label{2.6}u\frac{\partial u}{\partial x}+v\frac{\partial u}{\partial y}=u_e%
\frac{\partial u_e}{\partial x}+\nu \omega \frac \partial {\partial y}\left(
\beta \frac{\partial u}{\partial y}\right) +\gamma \frac 1{\rho _m}\sigma
(x,y)E_x(x,y) 
\end{equation}
with 
\begin{equation}
\label{2.7}\beta =1+\frac \epsilon \omega 
\end{equation}
being, in general, a function of $y$ through the dependence of the eddy
viscosity $\epsilon $ on the local velocity field. $\beta $ should be
obtained from a turbulence model.

To analyze the scaling solutions and for the numerical calculations in Sect.
3 we define a stream function $\psi $ and make the following change of
variables 
\begin{equation}
\label{2.8a}\eta =\left( \frac{u_e}{\nu \omega }\right) ^{\frac 12}\frac
y{\xi (x)}
\end{equation}
\begin{equation}
\label{2.8b}\psi =\left( u_e\nu \omega \right) ^{\frac 12}\xi (x)f(x,\eta )
\end{equation}
\begin{equation}
\label{2.8c}u=\frac{\partial \psi }{\partial y}\bigskip\ ,\bigskip\ v=-\frac{%
\partial \psi }{\partial x}
\end{equation}
The continuity equation (\ref{2.3a}) is automatically satisfied by (\ref
{2.8c}) and one is left with 
\begin{equation}
\label{2.9}\frac \partial {\partial \eta }\left( \beta \frac{\partial ^2f}{%
\partial \eta ^2}\right) +\xi \frac{\partial \xi }{\partial x}f\frac{%
\partial ^2f}{\partial \eta ^2}+\frac{\xi ^2}{u_e}\frac{\partial u_e}{%
\partial x}+\xi ^2\left( \frac{\partial ^2f}{\partial \eta ^2}\frac{\partial
f}{\partial x}-\frac{\partial f}{\partial \eta }\frac{\partial ^2f}{\partial
\eta \partial x}\right) =-\frac \gamma {u_e^2\rho _m}\xi ^2(x)\sigma (x,\eta
)E_x(x,\eta )
\end{equation}

\subsection{Scaling solutions}

We assume that the electric field to be created by a series of plate
electrodes along the z-direction, that is transversal to the fluid flow. For
this electrode geometry the mean electric field in the x-direction may be
parametrized by 
\begin{equation}
\label{2.10}\overline{E_x}=g(x)\frac{l(x)}{l^2(x)+y^2} 
\end{equation}
where $x$ and $y$ are the adimensional coordinates defined in (\ref{2.2a}). $%
E_0=\frac{g(x)}{l(x)}$ is the field at $y=0$, controlled by the potential
differences between the electrodes, and $l(x)$ is of the order of the
electrode spacing. For thin (laminar) boundary layers the field $E_x$ may
with good approximation be considered to be independent of $y$ throughout
the boundary layer thickness, as long as the appropriate charge density
profile is chosen (see below).

For the main application we are addressing, ionized air would be injected
through a slot near the leading edge of the airfoil, being then carried
along the airfoil surface by the flow. The steady-state charge distribution
in the boundary layer is obtained from the continuity equation 
\begin{equation}
\label{2.10a}\frac{\partial \sigma }{\partial x}u+\frac{\partial \sigma }{%
\partial y}v=j 
\end{equation}
$j$ being the source of electric charge. For a point source at the position (%
$x_0,y_0$), that is $j=c(x_0,y_0)\delta (x-x_0)\delta (y-y_0)$, the solution
is 
\begin{equation}
\label{2.10b}\sigma (x,y)=c(x_0,y_0)\delta \left( \psi (x,y)-\psi
(x_0,y_0)\right) \theta (x-x_0) 
\end{equation}
$\psi $ being the stream function defined before. Then, for a column of
ionized air injected at a backwards facing angle through a slot placed at $%
x_0$, behind the stagnation point, each point acts as a point source of
intensity proportional to the local fluid velocity. Furthermore the
intensity of the effective source is depleted up the column. Taking the
depletion effect into account, one obtains by integration of Eq.(\ref{2.10b}%
) 
\begin{equation}
\label{2.10c}\sigma (x,y)=\sigma _0\left( 1-d_1\psi (x,y)\right) \theta
\left( 1-d_1\psi (x,y)\right) \theta (x-x_0) 
\end{equation}
$\sigma _0$ is the injection intensity and $d_1$ characterizes the rate of
depletion. The conclusion is that the charge density is maximum at the
airfoil surface, decreasing to zero at a distance that depends on the fluid
dynamics and the injection regime. In numerical simulations one may easily
use the fairly accurate equation (\ref{2.10c}) for the charge density
profile. Here however, the dynamically-dependent charge density profile will
be parametrized by the simpler formula 
\begin{equation}
\label{2.11}\sigma (x,y)=\sigma _0\left( 1-\frac u{u_e}\right) 
\end{equation}

We now look for scaling solutions of (\ref{2.9}). A scaling solution is one
for which $f$ is only a function of $\eta $. Eq.(\ref{2.9}) becomes 
\begin{equation}
\label{2.12a}\left( \beta f^{^{\prime \prime }}\right) ^{^{\prime }}+\xi 
\stackrel{\bullet }{\xi }ff^{^{\prime \prime }}+\frac{\xi ^2}{u_e}\frac{%
\partial u_e}{\partial x}=-\frac \gamma {u_e\rho _m}\xi ^2(x)\sigma _0\left(
1-f^{^{\prime }}\right) g(x)\frac{l(x)}{u_el^2(x)+\omega \nu \xi ^2(x)\eta ^2%
}
\end{equation}
with boundary conditions 
\begin{equation}
\label{2.12b}f(0)=f^{^{\prime }}(0)=0\bigskip\ \bigskip\ f^{^{\prime
}}(\infty )=1
\end{equation}
where, for simplicity, we have denoted $f^{^{\prime }}\equiv \frac{\partial f%
}{\partial \eta }$ and $\stackrel{\bullet }{\xi }=\frac{\partial \xi }{%
\partial x}$.

Let the pressure be approximately constant for length scales $L$ of the
order of the airfoil, that is $\frac{\partial u_e}{\partial x}\approx 0.$
Let also $\beta $ be a constant. This is the case for the laminar part of
the boundary layer. Then the factorized nature of Eq.(\ref{2.12a}) implies
that solutions exist only if 
\begin{equation}
\label{2.13}\xi ^{-2}(x)=\frac 2{c_1}\frac{\stackrel{\bullet }{\xi }(x)}{\xi
(x)}=\frac 1{c_3}\frac{g(x)}{\xi (x)}=c_4l^{-2}(x)
\end{equation}
$c_1$, $c_3$ and $c_4$ being constants. Therefore 
\begin{equation}
\label{2.14}
\begin{array}{c}
\xi (x)=
\sqrt{c_1x+c_2} \\ g(x)=
\frac{c_3}{\xi (x)} \\ l(x)=\sqrt{c_4}\xi (x)
\end{array}
\end{equation}
There are two physically interesting situations. The one with $c_1\neq 0$ $%
c_2=0$ and the one with $c_1=0$ $c_2\neq 0$. The first one corresponds to a
boundary layer starting at $x=0$ and growing with $x^{\frac 12}$ and the
second to a constant thickness boundary layer. The first one corresponds to
an equation 
\begin{equation}
\label{2.15a}f^{^{\prime \prime \prime }}(\eta )+\frac 12f(\eta )f^{^{\prime
\prime }}(\eta )+\left( 1-f^{^{\prime }}(\eta )\right) \frac{\varphi _1}{%
\varphi _2^2+\eta ^2}=0
\end{equation}
with $c_1=\beta $ , $c_2=0$ , $\varphi _1=\frac{\gamma \sigma _0c_3\sqrt{c_4}%
}{u_e\beta \rho _m\nu \omega }$ , $\varphi _2=\sqrt{\frac{u_ec_4}{\nu \omega 
}}$ and the second to 
\begin{equation}
\label{2.15b}f^{^{\prime \prime \prime }}(\eta )+\left( 1-f^{^{\prime
}}(\eta )\right) \frac a{b^2+\eta ^2}=0
\end{equation}
with $c_1=0$ , $c_2\neq 0$ , $a=\frac{\gamma \sigma _0c_3\sqrt{c_4}}{%
u_e\beta \rho _m\nu \omega }$ , $b=\sqrt{\frac{u_ec_4}{\nu \omega }}$ .

In the first case one chooses $c_2=0$ to obtain a boundary layer starting at 
$x=0$. The scaling hypothesis requires then an electric field that is
singular at $x=0$, $y=0$ ($E_x\sim x^{-1}$). In any case this electric field
solution is not very interesting for our purposes because it leads to a
boundary layer growth of $x^{\frac 12}$, as in the free force Blasius
solution. Therefore it will be more interesting to consider a small field
free region in the leading edge of the airfoil and match the Blasius
solution there with the constant thickness solution of Eq.(\ref{2.15b}).

Gailitis and Lielausis\cite{Gailitis} have also obtained a theoretical
solution of constant thickness. However they consider a different force
field distribution and no dependence of the fluid charge density on the
boundary layer dynamics. Therefore their boundary layer profile has a very
different behavior.

The solution of Eq.(\ref{2.15b}) is easily obtained by numerical integration
(see below). Notice however that with the replacement 
\begin{equation}
\label{2.16}\phi (\eta )=1-f^{^{\prime }}(\eta ) 
\end{equation}
and choosing $c_4=\frac{\nu \omega }{u_e}$ , which is a simple rescaling of $%
\xi $, Eq.(\ref{2.15b}) becomes the zero-eigenvalue problem for a
Schr\"odinger equation in the potential $a/(1+\eta ^2)$, 
\begin{equation}
\label{2.17}-\phi ^{^{\prime \prime }}(\eta )+\frac a{1+\eta ^2}\phi (\eta
)=0 
\end{equation}
One may use the well-known WKB approximation to obtain 
\begin{equation}
\label{2.18}f^{^{\prime }}(\eta )=1-\frac{\left( 1+\eta ^2\right) ^{\frac 14}%
}{\left( \eta +\sqrt{1+\eta ^2}\right) ^{\sqrt{a}}} 
\end{equation}
Eq.(\ref{2.18}) is a very good approximation to the exact solution for $%
a\geq 1$ (see Fig. 2). Fig.3 shows the effective boundary layer thickness as
a function of $a$. The effective boundary layer thickness $\delta ^{*}$ is
defined here as the value of $\eta $ at which the velocity $u$ reaches 0.95
of its asymptotic value $u_e$. A very fast thinning of the boundary layer is
obtained (several orders of magnitude) for a relatively short range of the $%
a $ parameter. Fig.3 shows the variation of $\delta ^{*}$ for small $a$. For
large $a$ (and small $\delta ^{*}$) one has the asymptotic formula%
$$
\delta ^{*}\simeq \frac{2.9957}{\sqrt{a}} 
$$
which is obtained from Eq.(\ref{2.18}).

If the longitudinal electric field $E_x$ is assumed to be a constant ($E_0$)
throughout the boundary layer thickness, with the same charge profile, the
solution is even simpler, namely 
\begin{equation}
\label{2.19}f^{^{\prime }}(\eta )=1-e^{-\eta \sqrt{h}}
\end{equation}
with $\xi =\sqrt{c_2}$ and 
\begin{equation}
\label{2.20}h=\frac{\gamma c_2\sigma _0E_0}{\beta u_e^2\rho _m}
\end{equation}
Again, since $\xi $ is a constant, this is not fully realistic because it
leads to a constant thickness boundary layer.

For reference values of the physical quantities in Eqs.(\ref{2.2a}-\ref{2.2c}%
) we take 
\begin{equation}
\label{2.21}
\begin{array}{c}
U_r=100 
\textnormal{ m s}^{-1} \\ L_r=1 
\textnormal{ m} \\ \delta _r=10^{-3} 
\textnormal{ m} \\ \rho _r=1.2 
\textnormal{ Kg m}^{-3} \\ E_r=500 
\textnormal{ V cm}^{-1} \\ \sigma _r=15 
\textnormal{ }\mu \textnormal{C cm}^{-3} \\ \nu _r=1.5\times 10^{-5}\textnormal{ m}^2\
textnormal{ s}^{-1} 
\end{array}
\end{equation}
For these reference values, the adimensional constants $\omega $ and $\gamma 
$ defined after Eq.(\ref{2.3c}) are 
\begin{equation}
\label{2.22}
\begin{array}{c}
\omega =0.15 \\ 
\gamma =62.499 
\end{array}
\end{equation}

For comparison we mention that in the classical force-free Blasius solution,
and for these reference parameters, the $y-$coordinate $y^{*}$ corresponding
to $\delta ^{*}$ (that is, the point at which $\frac u{u_e}=0.95$) is 
\begin{equation}
\label{2.23}y^{*}=1.55\times 10^{-3}\sqrt{x} 
\end{equation}
Stability of a laminar boundary layer cannot safely be guaranteed for local
Reynold numbers greater than about $10^3$. Therefore requiring 
\begin{equation}
\label{2.24}R_S=\frac{\widetilde{u_e}\widetilde{y}^{*}}{\widetilde{\nu }}%
\simeq 10^3 
\end{equation}
one obtains, for the reference parameters, $\widetilde{y}^{*}\simeq 0.15$
mm. Using (\ref{2.21}) the conclusion is that, for these parameters, the
laminar part of a force-free boundary layer is only of the order of $1$ cm,
just a tiny portion of a typical wing.

Now we use the scaling solutions (\ref{2.19}) and (\ref{2.18}) to obtain an
estimate of the effects of a longitudinal electric field. For the constant
field case (\ref{2.19}) from%
$$
f^{^{\prime }}(\delta ^{*})=1-e^{-\delta ^{*}\sqrt{h}}=0.95 
$$
and%
$$
y^{*}=\delta ^{*}\sqrt{c_2}\sqrt{\frac{\nu \omega }{u_e}}=0.15 
$$
using (\ref{2.20}) one obtains%
$$
\sigma _0=0.957 
$$
That is, to insure a constant thickness boundary layer with local Reynolds
number $R_S=10^3$ (at the point where $\frac u{u_e}=0.95$), one needs a
charge density $\widetilde{\sigma }_0$ at $y=0$, in physical units (and for
the reference values of the kinematical parameters)%
$$
\widetilde{\sigma }_0=\sigma _0\sigma _r=14.36\textnormal{ }\mu \textnormal{C cm}^{-3} 
$$

For the variable field case (\ref{2.18}) the estimate depends on the
separation of the electrodes. Taking $l(x)=10$, that is an electrode
separation of the order of one centimeter, and the references values for all
quantities except for the charge density (namely $E_0=\frac{g(x)}{l(x)}=1$, $%
u_e=1$, etc.) one obtains $c_4=0.15$, $g(x)=10$, $c_3=g(x)\xi (x)=258.2$, $%
\sqrt{c_2}=\xi (x)=25.8$, and requiring%
$$
\begin{array}{c}
y^{*}=\delta ^{*} 
\sqrt{c_2}\sqrt{\frac{\nu \omega }{u_e}}=0.15 \\ f^{^{\prime }}(\delta
^{*})=0.95 
\end{array}
$$
one finally obtains $a=39887.77$ leading to%
$$
\sigma _0=0.957 
$$
the same estimate as above. The large value of $a$ that is obtained shows
that the WKB expression (\ref{2.18}) is a good approximation for physically
interesting parameter values. On the other hand the fact that the same
charge density estimate is obtained both in the constant-field and the
variable-field cases, shows that it is realistic to consider the field as
approximately constant throughout the laminar boundary layer thickness, as
long as a variable charge profile (\ref{2.10c}) or (\ref{2.11}) is used.

The above estimates were obtained using the reference values for the
kinematic variables. For other values we have the following designing
formula (in normalized units) 
\begin{equation}
\label{2.30}\sigma _0E_0=0.957\frac{u_e^3\rho _m}{10^{-6}R_S^2\nu }
\end{equation}

\section{Numerical results}

For the numerical solution of Eq.(\ref{2.9}), with $\sigma $ given by Eq.(%
\ref{2.11}), we use an implicit finite-difference technique (\cite{Blottner}
- \cite{Hamilton}). Define $F(x,\eta )$ by 
\begin{equation}
\label{3.1}F(x,\eta )=\frac{\partial f}{\partial \eta }
\end{equation}
and 
\begin{equation}
\label{3.2}
\begin{array}{c}
a_1=\frac 1\beta \left( 
\frac{\partial \beta }{\partial \eta }+\xi \frac{\partial \xi }{\partial x}%
f+\xi ^2\frac{\partial f}{\partial x}\right)  \\ a_2=-\frac \gamma {\beta
u_e^2\rho _m}\xi ^2E_x\sigma _0 \\ 
a_3=-
\frac{\xi ^2}\beta F \\ a_4=\frac \gamma {u_e^2\rho _m\beta }\xi ^2E_x\sigma
_0+\frac{\xi ^2}{u_e\beta }\frac{\partial u_e}{\partial x}
\end{array}
\end{equation}
Then Eq.(\ref{2.9}) becomes 
\begin{equation}
\label{3.3}\frac{\partial ^2F}{\partial \eta ^2}+a_1\frac{\partial F}{%
\partial \eta }+a_2F+a_3\frac{\partial F}{\partial x}+a_4=0
\end{equation}
The derivatives are replaced by finite-difference quotients with a variable
grid spacing concentrated near $\eta =0$, where $F$ changes more rapidly.
Let $k>1$ be the ratio between two successive grid spacings in the $\eta -$%
direction.%
$$
k=\frac{\eta _{i+1}-\eta _i}{\eta _i-\eta _{i-1}} 
$$
Then%
$$
\begin{array}{c}
\left( 
\frac{\partial ^2F}{\partial \eta ^2}\right) _{i+1,j}=2\frac{%
F_{i+1,j+1}+kF_{i+1,j-1}-(1+k)F_{i+1,j}}{\Delta _2} \\ \left( 
\frac{\partial F}{\partial \eta }\right) _{i+1,j}=\frac{%
F_{i+1,j+1}-k^2F_{i+1,j-1}-(1-k^2)F_{i+1,j}}{\Delta _1} \\ \left( 
\frac{\partial F}{\partial x}\right) _{i+1,j}=\frac{F_{i+1,j}-F_{i,j}}{%
\Delta x} \\ \Delta _1=\eta _{j+1}-\eta _j+k^2(\eta _j-\eta _{j-1}) \\ 
\Delta _2=(\eta _{j+1}-\eta _j)^2+k(\eta _j-\eta _{j-1})^2
\end{array}
$$
Substitution in Eq.(\ref{3.3}) yields 
\begin{equation}
\label{3.6}A_jF_{i+1,j+1}+B_jF_{i+1,j}+D_jF_{i+1,j-1}+G_j=0
\end{equation}
with%
$$
\begin{array}{c}
A_j=\frac 2{\Delta _2}+
\frac{a_1}{\Delta _1} \\ B_j=
\frac{-2(1+k)}{\Delta _2}-\frac{a_1(1-k^2)}{\Delta _1}+a_2+\frac{a_3}{\Delta
x} \\ D_j=
\frac{2k}{\Delta _2}-\frac{a_1k^2}{\Delta _1} \\ G_j=a_4-a_3\frac{F_{i,j}}{%
\Delta x}
\end{array}
$$
The boundary conditions at $\eta =0$ and $\eta \rightarrow \infty $ are known%
$$
\begin{array}{c}
f(i,1)=F(i,1)=0 \\ 
F(i,N)=1
\end{array}
$$
where $N$ is the largest label of the grid, in the $\eta -$coordinate,
chosen to be sufficiently large.

Because of the tridiagonal nature of (\ref{3.6}) the solution in the line $%
i+1$ is obtained by the two-sweep method, the recursion relations being%
$$
\begin{array}{c}
F_{i+1,j}=\alpha _jF_{i+1,j+1}+\beta _j \\ 
\alpha _j=- 
\frac{A_j}{B_j+D_j\alpha _{j-1}} \\ \beta _j=-\frac{G_j+D_j\beta _{j-1}}{%
B_j+D_j\alpha _{j-1}} 
\end{array}
$$
with $\alpha _1=0$ and $\beta _1=0$.

To start the integration process there are basically two methods. In the
first the integration is performed from left to right in the $x-$coordinate
with the grid extended to the left of the airfoil, where the flow is known.
With the solution known in the line $i$, the coefficients $A_j$ to $G_j$ for
Eq.(\ref{3.6}) are computed at the point ($i,j$). Notice that $f(i,j)$ is
obtained by integration of the solution $F$.%
$$
f(i,\eta )=\int_0^\eta F(i,\zeta )d\zeta 
$$
The integration now proceeds along the lines, from left to right. After a
complete pass the process is restarted using now for the calculation of the
coefficients $A_j$, $B_j$, $D_j$ and $G_j$ the old values of $F$ at $(i+1,j)$%
. The process is repeated several times until the solution stabilizes.

In the second method, which is the one we actually use, the integration
process starts from an approximate solution. The scaling solutions derived
in Sect.2 are particularly useful for this purpose.

For our calculations we considered an electric field parametrized as in Eq.(%
\ref{2.10}), namely%
$$
E_x=E_0\frac{\frac{u_el^2}{\nu \omega }}{\frac{u_el^2}{\nu \omega }+\xi
^2(x)\eta ^2} 
$$
with $\frac{u_el^2}\omega =666.66$ which corresponds to $l=10$, $u_e=1$ and $%
\nu \omega =0.15$. Notice that for these parameters, as pointed out in
Sect.2, the electric field has only a small variation throughout the
boundary layer region. For the scaling function we take $\xi (x)=\sqrt{x}$
and consider $\beta =1$. Then all results depend only on the variable $S$%
$$
S=\frac 1{62.499}\frac \gamma {u_e^2\rho _m}\sigma _0E_0 
$$
($S=1$ when all quantities take the reference values).

In Fig.4 we show a contour plot of the numerical solution for $f^{^{\prime
}}(x,\eta )$ (=$\frac u{u_e}$) when $S=0.6$. From the $x-$dependence of the
numerical solutions we may compute the effect of the electric field in
extending the laminar part of the boundary layer. By defining, as in Sect.2,
the length of the laminar part as the $x-$coordinate corresponding to a
local Reynolds number of $10^3$ and denoting by $x_0$ $(\frac u{u_e}=0.95)$
the force-free value we have obtained for the ratio%
$$
R=\frac x{x_0} 
$$
the results shown in Fig.5. For $S=0$ we obtain the Blasius solution and as
we approach $S=0.957$, corresponding to the scaling solution, the ratio
diverges. The matching of the results in the force-free and scaling limits
is a good check of the numerical algorithm. A clear indication of the
results in Fig.5 is that not much improvement is obtained unless one is able
to obtain ionization charge densities of the order of the reference value $%
\sigma _r$.

\section{Discussion and conclusions}

\# In this paper we have concentrated on controlling the profile of the
boundary layer. The profile has a direct effect on the laminar or turbulent
nature of the flow which, in a simplified manner, we estimated by a local
Reynolds number (\ref{2.24}) defined as a function of the effective
thickness. Another relevant aspect, of course, is the active control of the
transition instabilities that can be achieved by electromagnetic body forces
on the charged fluid.

Turbulence and transition to turbulence are three-dimensional phenomena.
However, for the large scale small amplitude (Tollmien-Schlichting)
fluctuations, that first appear in the transition region, a two-dimensional
model is a reasonable approximation. In Eqs.(\ref{2.1a}-\ref{2.1c}) we make
as before the change of variables (\ref{2.2a}-\ref{2.2c}), neglect terms of
order $\frac 1{R_L}$ , $\frac{\delta _r^2}{L_r^2}$ and $\frac{\delta _r}{L_r}
$ and split the velocity and electric fields into%
$$
\begin{array}{c}
u= 
\overline{u}+u^{^{\prime }} \\ v= 
\overline{v}+v^{^{\prime }} \\ E_x=\overline{E_x}+E_x^{^{\prime }} 
\end{array}
$$
where $\overline{u}$ , $\overline{v}$ , $\overline{E_x}$ are the
steady-state solutions and $u^{^{\prime }}$ , $v^{^{\prime }}$ , $%
E_x^{^{\prime }}$ the time-dependent components. Because of the continuity
equation%
$$
\frac{\partial u^{^{\prime }}}{\partial x}+\frac{\partial v^{^{\prime }}}{%
\partial y}=0 
$$
we may define a fluctuation stream function $\chi $%
$$
u^{^{\prime }}=\frac{\partial \chi }{\partial y}\bigskip\ ,\bigskip\
v^{^{\prime }}=-\frac{\partial \chi }{\partial x} 
$$
Now we assume the fluctuation to be a (small-amplitude) wave-like function
of $x$, $y$ and $t$%
$$
\chi (x,y,t)=F(y)e^{i(\alpha x-\theta t)} 
$$
The imaginary parts of $\theta $ and $\alpha $ control, respectively, the
growth rates of temporal and spatial fluctuations. The (modulation) electric
field is assumed to have a similar form%
$$
E_x^{^{\prime }}=Ee^{i(\alpha x-\theta t)} 
$$
One now obtains%
$$
\left( i\overline{u}\alpha +\frac{\partial \overline{u}}{\partial x}-i\theta
\right) \frac{\partial F}{\partial y}-i\alpha \frac{\partial \overline{u}}{%
\partial y}F+\overline{v}\frac{\partial ^2F}{\partial y^2}=\nu \omega \frac{%
\partial ^3F}{\partial y^3}+\frac \gamma {\rho _m}\sigma E 
$$
The conclusion is that a space-time modulation of the electric field, with
the appropriate phase, is equivalent to an effective viscous damping effect
which delays the growth of the transition region instability. For this to be
effective one needs to detect the phase of the wave instabilities by
electromagnetic probes. Absolute synchronization of the feedback electric
modulation is however not so critical as in acoustic noise cancelation,
because here the objective is only to obtain an effective damping effect.
The simplified treatment of the transition instabilities is justified by the
fact that it is only for the small amplitude large scale fluctuations that
one may hope to be able to detect the phase with some reasonable accuracy.

\# The kinematic reference parameters defined in (\ref{2.21}) correspond to
typical aviation conditions. The conclusion, both from the scaling solutions
in Sect.2 and the numerical results in Sect.3, is that, to obtain a
significant controlling effect on the boundary layer by this method, the
charge density $\sigma _0$ (at $y=0$) should be of the order of the
reference charge density ($\sigma _r=15\mu $C cm$^{-3}$). This charge
density corresponds to about 50 times the ion concentration a few
centimeters away from the emitter of a commercial table-top negative corona
discharge air purifier with a power of less than 6 watts. Therefore, it
seems technically feasible to achieve a significant boundary layer control
by this method. Another possibility would be to use, instead of air, some
other easier to ionize gas. This could then be partially recovered and
recycled by suction.

\# As explained in the introduction and because of the perturbation induced
by the injection method, it seems advisable to use this method in
conjunction with suction and passive control in the rear part of the
airfoil. Even if a fully laminar boundary layer may never be completely
achieved, just remember that any small improvement becomes, in the long run,
quite significant in terms of fuel consumption.

\# The formula (\ref{2.30}), derived from the scaling solutions, provides
rough design estimates. Better control over design parameters we hope to
obtain from the experimental work.

\section{Figure captions}

Fig.1 Airfoil transversal cut showing ionized air injection, suction pump
and plate electrodes.

Fig.2 Exact ($-$) and approximate ($\cdots $) constant thickness scaling
solution $f^{^{\prime }}(\eta )$.

Fig.3 Effective boundary layer thickness $\delta ^{*}$ $\left( f^{^{\prime
}}(\delta ^{*})=\frac u{u_e}=0.95\right) $ for the constant thickness
scaling solution.

Fig.4 Contour plot of $f^{^{\prime }}(x,\eta )$ for $S=0.6$.

Fig.5 Ratio of boundary layer laminar regions with and without electric
field control.

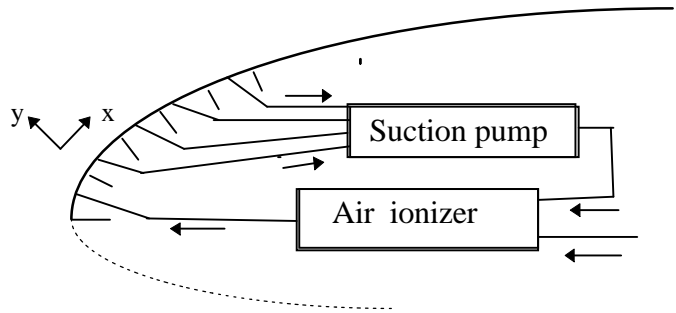
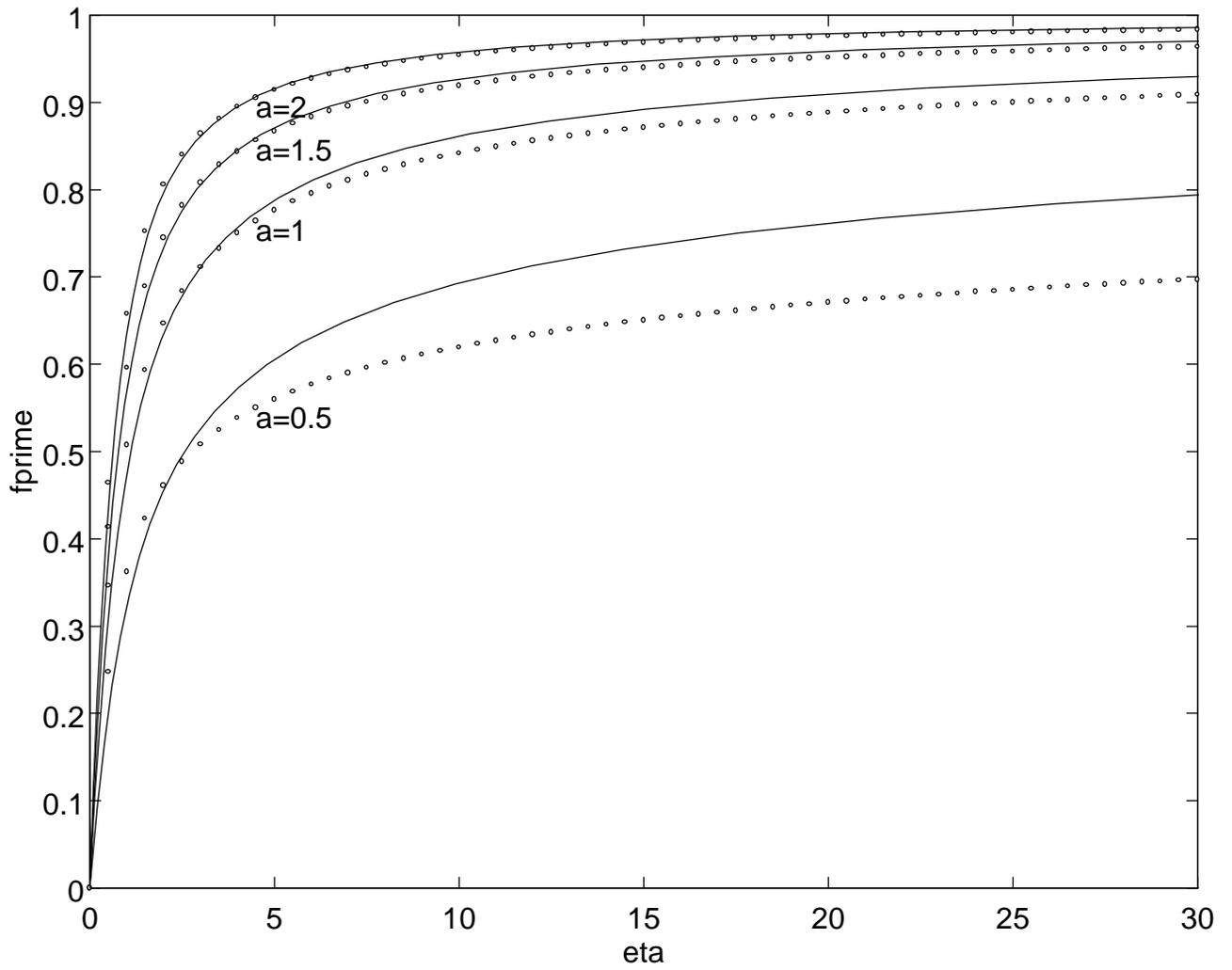
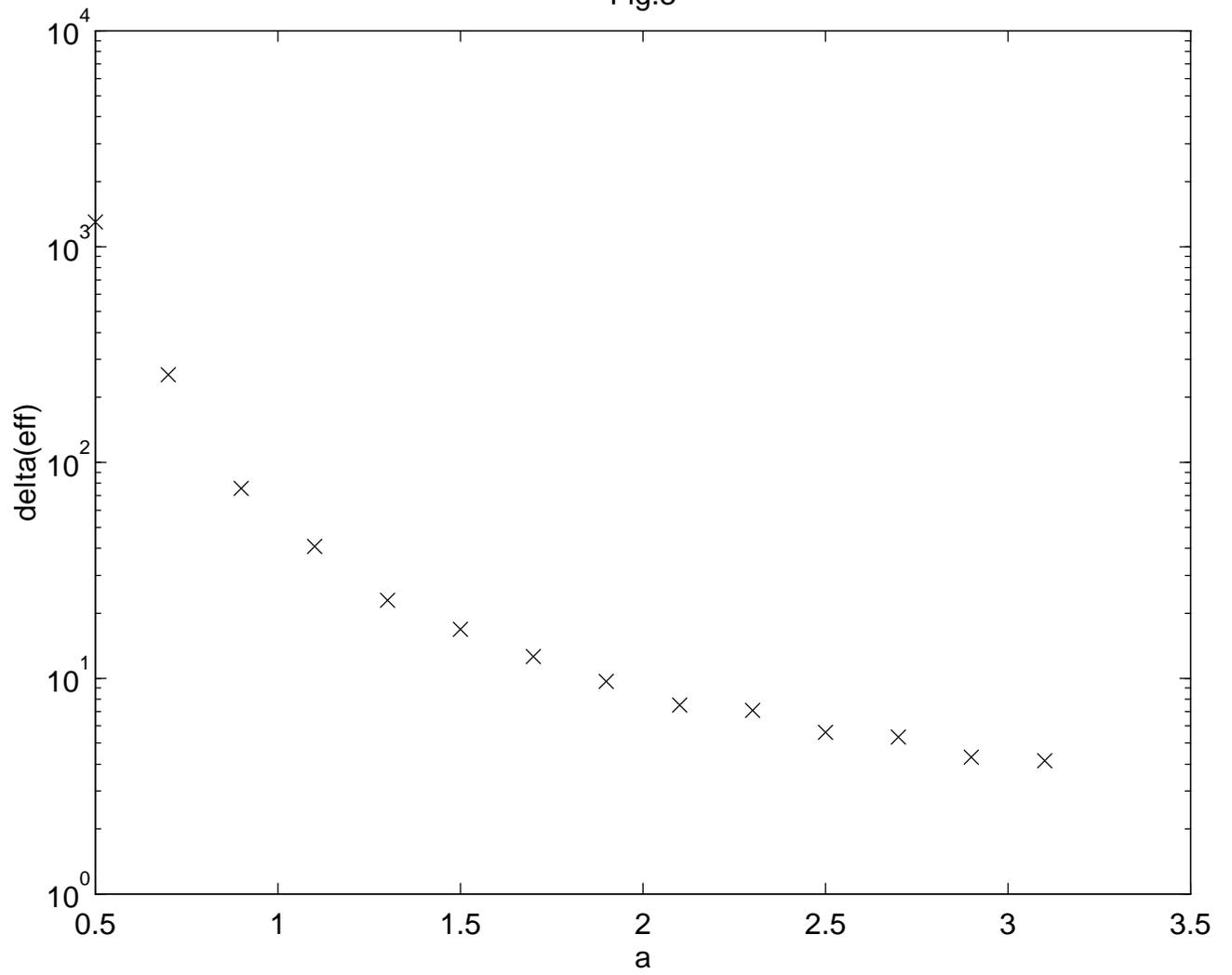
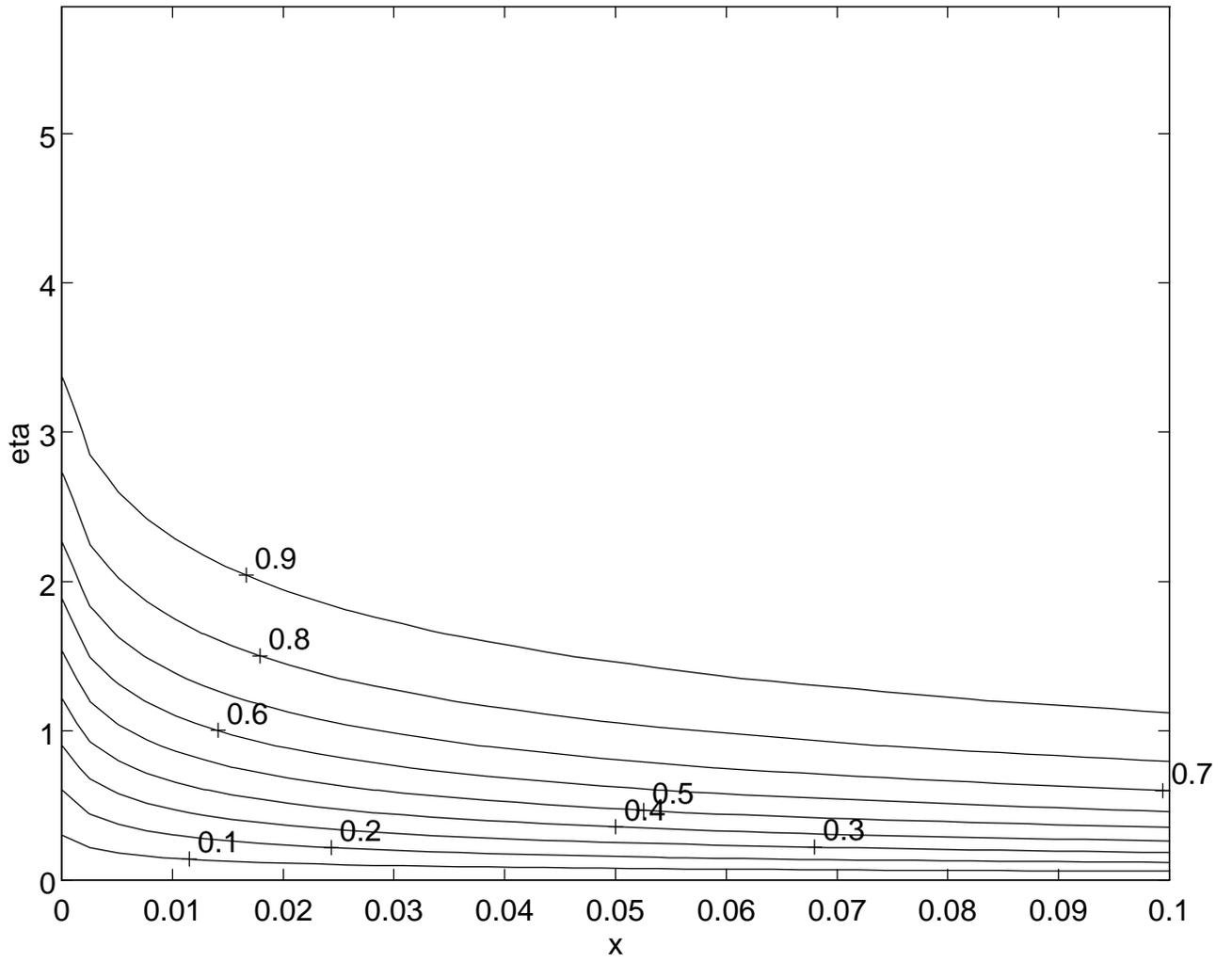
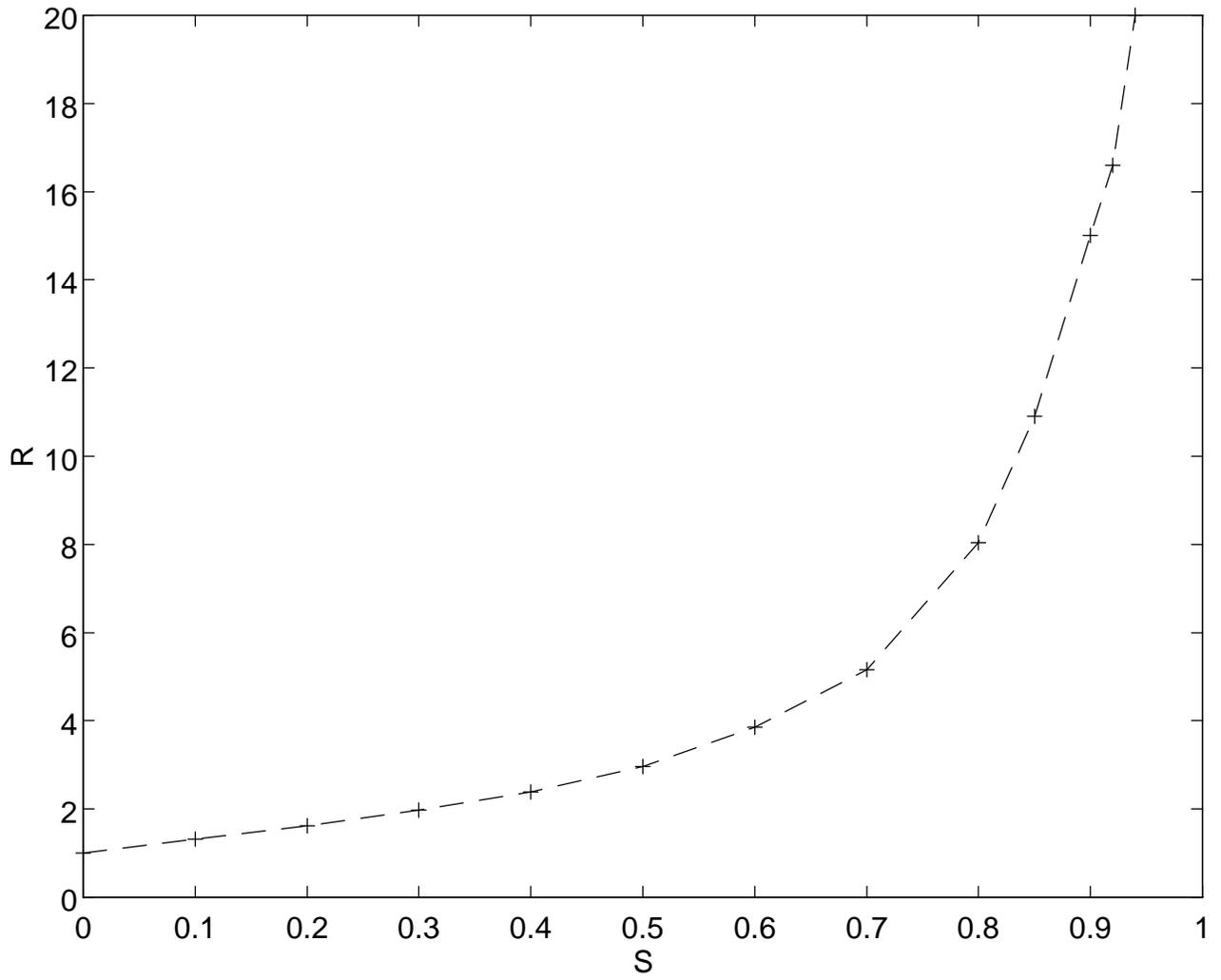
\end{document}